\documentclass[twocolumn,showpacs,preprintnumbers,amsmath,amssymb,tightenlines,epsfig]{revtex4}

\usepackage{bm}
\usepackage{graphicx}

\usepackage{graphicx}
\usepackage{dcolumn}
\usepackage{bm}

\newcommand{\ket}[1]{|#1\rangle}

\newcommand{\beq}{\begin{equation}}
\newcommand{\eeq}{\end{equation}}

\newcommand{\bml}{\begin{mathletters}}
\newcommand{\eml}{\end{mathletters}}
\newcommand{\commentout}[1]{{}}

\normalsize

\begin{document}

\title{Effective one-component description of two-component Bose-Einstein condensate dynamics}

\author{Zachary Dutton and Charles W. Clark}
\affiliation{National Institute of Standards Technology, Electron
and Optical Division, Gaithersburg MD 20899-8410}
\date{\today}

\begin{abstract}

We investigate dynamics in two-component Bose-Einstein condensates
in the context of coupled Gross-Pitaevskii equations and derive
results for the evolution of the total density fluctuations. Using
these results, we show how, in many cases of interest, the
dynamics can be accurately described with an effective
one-component Gross-Pitaevskii equation for one of the components,
with the trap and interaction coefficients determined by the
relative differences in the scattering lengths. We discuss the
model in various regimes, where it predicts breathing excitations,
and the formation of vector solitons.   An effective nonlinear
evolution is predicted for some cases of current experimental
interest.  We then apply the model to construct quasi-stationary
states of two-component condensates.

\end{abstract}

\pacs{03.67.-a, 03.75.Lm, 03.75.Mn, 42.50.Gy}

\maketitle

\section{\label{sec:intro} Introduction}

One of the most intriguing aspects of recently produced atomic
Bose-Einstein condensates \cite{BEC} is the ease with which two (or
more) internal levels can be populated, in effect forming a spin-1/2
(or spin-1, etc.) system.  In particular, phase separation
\cite{phaseSepExp, phaseSepTh}, vector solitons
\cite{twoCompSolitonsBEC}, two-component vortices
\cite{twoCompVortex,twoCompVortexLatt,vortexStorage,twoCompVortexStab},
spin waves \cite{spinWaves}, breathe-together solutions
\cite{dynamics,breatheTogether} and more general issues regarding
the dynamics \cite{multipleComponentJILA,symBreaking,dynamics} and
stability \cite{twoCompStab} in these systems have recently been
investigated. Two-component BECs hold promise for a number of
applications, including coherent storage and processing of optical
fields \cite{processing} using stopped light techniques
\cite{stoppedLight}.

The dynamics of two-component BECs is in general a difficult problem
to study analytically, and much of the recent activity has focused
on particular situations and thus made specific assumptions.  For
example, the relative interaction strengths (via the binary
collision scattering lengths) for inter- and intra-component
collisions are sometimes presumed to be equal, whereas in practice
these scattering lengths vary and depend on the atom and the
particular levels utilized.  Furthermore, these can be
experimentally varied with Feshbach resonances \cite{feshbach}. In
other studies it is assumed there is little or no spatial variation
in the relative density between the two components, restricting the
applicability of these calculations to many cases of interest.  For
example, in stopped light experiments, a highly spatially dependent
superposition of the two internal atomic states are generated by
light pulses, and the dynamics of this inhomogenous superposition is
essential in determining the information storage and processing
capabilities of the system \cite{processing}.

The need for a simple understanding of dynamics in two-component
BECs free of such assumptions is the motivation for this work.  We
specifically consider the experimentally relevant situation of a
single-component BEC (in an internal state labelled $\ket{1}$) in
its ground state, whose constituent atoms are suddenly put into
spatially dependent superposition of two internal levels ($\ket{1}$
and $\ket{2}$), with the $\ket{2}$ condensate occupying a region
inside the larger $\ket{1}$ condensate.  Such a situation occurs
with stopped light pulses \cite{processing} or spatially dependent
Raman pulses. When such processes occur fast compared to the atomic
dynamics (millisecond timescales), the resulting two-component BEC
will initially have the same density profile as the original
single-component BEC. However, this is not a stationary state and
will evolve. The evolution can be modelled with coupled
Gross-Pitaevskii (GP) equations, where each component is described
with a single macroscopic wave functions $\psi_1$ and $\psi_2$.

We find, remarkably, that the evolution of $\psi_2$ can often be
described with an effective single-component GP equation, with the
trapping potential and interaction coefficients renormalized by the
fractional difference in scattering lengths.  Those these
differences are generally small in practice, they end up dominating
the motion of the wave functions. Depending on their sign and
magnitude, our model predicts both trapping and repulsive effective
potentials, as well as positive (repelling) or negative (attractive)
effective scattering lengths for the $\ket{2}$ condensate.

The reduction to a single-component picture is accomplished by
observing that the fluctuations of the total density are smaller and
more quickly varying than the evolution of each individual
component. We find an equation of motion for the total density
fluctuations and find that, within certain limits, these
fluctuations are described with a simple expression which we can
plug into the equation of motion for $\psi_2$ and derive our
effective one-component description.  While we will perform our
calculations with one-dimensional equations for computational
simplicity, the results are equally valid and applicable in full
three-dimensional geometries.

Using our model, much of the vast existing literature on
single-component BEC dynamics can be easily be applied to predict
analogous behavior in two-component systems. To demonstrate this
applicability, and test the accuracy of our model, we present
calculations in several parameter regimes.  For certain relative
scattering lengths we get a repulsive effective nonlinearity in a
trap. In this case a simple breathing motion occurs and the existing
analytic results on collective excitations in single-component BECs
\cite{oneCompExcitations} can be applied. In other cases, we get an
attractive nonlinearity, in which case phase separation occurs and
vector solitons form and propagate. Our model again allows a mapping
onto existing literature on the formation of soliton trains in
single component BECs \cite{solitonTrainExp,solitonTrainTh} which
should improve abilities to design and analyze experiments to
observe vector solitons, which have hitherto not been ovserved. We
also note that the levels $\ket{F=1,M_F=-1}$ and $\ket{F=2,M_F=+1}$
in ${}^{87}$Rb, used in many present day experiments
\cite{multipleComponentJILA,twoCompVortex}, give rise to a vanishing
effective nonlinearity, allowing us predict the evolution with a
linear Schroedinger equation.  This should prove especially powerful
in two and three dimensional cases, where it is computationally
expensive to solve the full nonlinear differential equations.
Finally, we use our model to predict the existence of solutions
whereby two overlapping condensates (each with arbitrary number)
remain nearly stationary for very long times.  These solutions can
be thought of as generalizing previously discussed breathe-together
solutions \cite{dynamics,breatheTogether}.  These solutions are
relevant to observing spin squeezing \cite{spinSqueezing} and
coherent optical storage \cite{processing} in these systems.

\section{\label{sec:derivation} Derivation of the one-component model}

\subsection{\label{subsec:description} Description of the system}

The coupled GP equations governing the evolution are

\begin{eqnarray}
\label{eq:GP2comp1} i \hbar \frac{\partial \psi_1 }{\partial t} & =
& \bigg[-\frac{\hbar^2}{2m}\frac{\partial^2}{\partial z^2}
\nonumber \\
& &  + V(z) + U_{11} | \psi_1 |^2   +  U_{12}| \psi_2 |^2 \bigg] \psi_1 \\
\label{eq:GP2comp2} i \hbar \frac{\partial \psi_2}{\partial t} & = &
\bigg[-\frac{\hbar^2}{2m}\frac{\partial^2}{\partial z^2} \nonumber
\\ & & + V(z) + U_{22} |\psi_2|^2 +  U_{12}| \psi_1 |^2 \bigg]
\psi_2,
\end{eqnarray}

\noindent where $m$ is the mass of the atoms and the harmonic
external trapping potential $V(z) = \frac{1}{2} m {\omega_z}^2 z^2$
is assumed to be equal for each component (which is true in
particular sub-levels for magnetic traps
\cite{multipleComponentJILA} for and for all sub-levels in far
off-resonant optical traps \cite{opticalTrap}). Atom-atom
interactions are characterized by the $U_{ij} = 4 \pi N \hbar^2
a_{ij}/m A$, where $N$ is the total number of condensate atoms,
$a_{ij}$ are the s-wave scattering lengths for binary collisions and
between atoms in internal states $|i \rangle$ and $|j \rangle$. We
are only accounting for dynamics in the the $z$ dimension, which is
valid in an elongated trap geometry where the significant dynamics
occur primarily in this dimension \cite{sound}, and so choose an
effective transverse area $A$ to give the correct nonlinearity . We
will show how the relative difference in the scattering lengths are
key parameters in the evolution and so define $\delta_c \equiv
(a_{12}-a_{11})/a_{11}$ and $\delta_2 \equiv
(a_{22}-a_{11})/a_{11}$.

\begin{figure}
\includegraphics[width=\columnwidth]{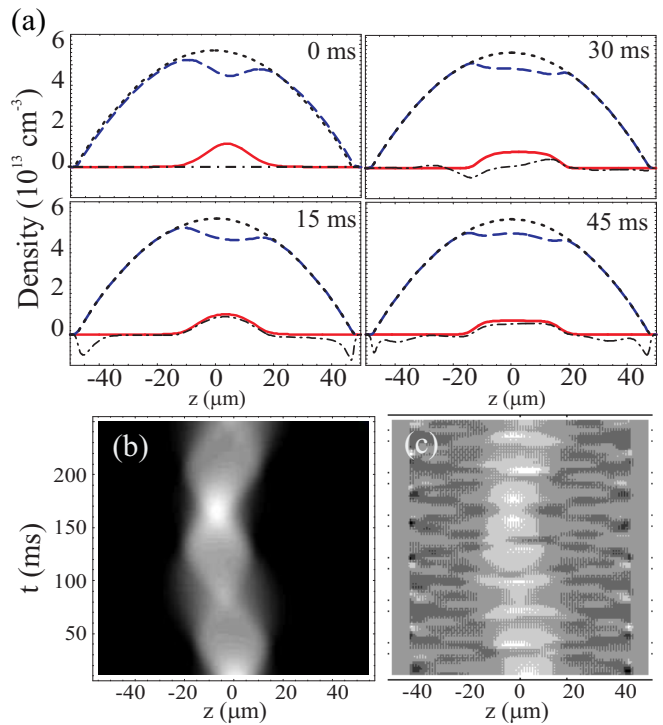}
\caption{\label{fig:breathEx}\textbf{(a)}  The densities $N
|\psi_2|^2$ (solid curve, red online) and $N |\psi_1|^2$ (dashed
curve, blue online), and total density $N \rho$ (dotted, black
curve) at the times indicated.  The initial total density $\rho_0$
corresponds to the ground state of a pure $\ket{1}$ condensate.
The dot-dashed curve shows the fluctuations of the total density
$\delta \rho$ scaled by $-1/\delta_c$. \textbf{(b)} The time
evolution of the density profile $|\psi_2|^2$ in this case.
\textbf{(c)} The time evolution of $\delta \rho$.}
\end{figure}

For concreteness, we consider a condensate with $N=2.0 \times 10^6$
atoms in a trap with frequency $\omega_z=(2 \pi)~21~\mathrm{Hz}$.
The area $A=\pi(4.2~ \mu \mathrm{m})^2$ is chosen such that the
initial density in the center corresponds to a ground state BEC in a
cylindrically symmetric trap with a transverse frequency $\omega_r=8
\omega_z$. The density profile of this initial state, labelled
$\rho_0$, is plotted as the dotted curve in the first panel of
Fig.~\ref{fig:breathEx}(a).  We then assume the BEC is put into a
spatial superposition of two states with a Gaussian-shaped and
slightly off-center wavefunction $\psi_2$, inside the larger
$\ket{1}$ condensate [again see the first panel of
Fig.~\ref{fig:breathEx}(a)].  The total density still given by
$\rho_0$ but, because of the spin excitation, this is no longer a
stationary state.

The resulting two-component evolution is presented in
Fig.~\ref{fig:breathEx}(a). Here we have chosen $\delta_2=-0.03$ and
$\delta_c=-0.03$.  As we observe in the figure, each component
undergoes non-trivial evolution of its density profile while the
overall density profile (which we define to be $\rho\equiv|\psi_1|^2
+ |\psi_2|^2$) undergoes only very small fluctuations. To get a
sense of the shape and magnitude of these small fluctuations, the
dot-dashed curves show the deviations from the original density
profile, $\delta \rho \equiv \rho - \rho_0$, blown up by the factor
$-1/\delta_c \approx 33$ (the reason for this particular case will
become apparent later). One sees that at 15~ms a slight fluctuation
in the total density has appeared in the region occupied by
$\ket{2}$ as well as near the condensate edges (in such a way that
the total atom number is conserved). Figure~\ref{fig:breathEx}(b)
gives an overview of the evolution of the density profile
$|\psi_2|^2$ and one sees that there is a breathing like behavior,
with the width becoming larger and then smaller, as well as a small
dipole sloshing due to its original offset from the trap center.
Figure~\ref{fig:breathEx}(c) then shows the evolution of the density
fluctuations $\delta \rho$ and one observes two important features.
First there is a pair of density perturbations travelling back and
forth across the $\ket{1}$ BEC (and reflecting at its boundaries),
which, as we will see below, are travelling at the sound speed
determined by the total BEC density. In the panels of
Fig.~\ref{fig:breathEx}(a), showing 15~ms and 45~ms, these waves are
near the BEC boundary whereas at 30~ms they are crossing in the
region occupied by $|\psi_2|^2$. Second, there is a part which
appears to be closely mimicking the much more slowly evolving
profile of $|\psi_2|^2$ (as also seen in the
Fig.~\ref{fig:breathEx}(a) plots).

\subsection{\label{subsec:densityFluctuations} Calculation of density
fluctuations}

We now analytically investigate the coupled GP equations
(\ref{eq:GP2comp1}-\ref{eq:GP2comp2}) to learn how this particular
pattern for the total density fluctuations arises.  Our strategy
will be to eliminate the wavefunction $\psi_1$ in favor of a
hydrodynamic description of the total density $\rho$ and total
velocity field $v_c=(v_1 |\psi_1|^2 + v_2 |\psi_2|^2)/\rho$, where
$v_i=(\hbar/m) \phi_i'$ are the velocity fields of each component
(the $\phi_i'$ denote the gradients of the phases of the
wavefunctions $\psi_i$; we will use the prime symbol to indicate
$\partial/\partial z$). We then linearize in the small density
fluctuations $\delta \rho$ and consequently small velocity field
$v_c$. Our observation that $\delta \rho$ evolves on a fast time
scale relative to $\psi_2$ supplies an obvious separation of time
scales in the problem and allows us to then solve for the evolution
$\delta \rho$ assuming $\psi_2$ is static on this time scale.

For convenience we define the relative density in $\ket{2}$ to be
$f=|\psi_2|^2/\rho$, so $1-f=|\psi_1|^2/\rho$.  After some
lengthy, but straightforward, algebra one can show that
Eqs.~(\ref{eq:GP2comp1}-\ref{eq:GP2comp2}) imply

\begin{eqnarray}
\label{eq:hydroEvol1} \dot{\rho} & = & - \rho v_c' - \rho' v_c, \\
\label{eq:hydroEvol2}\dot{v_c} & = & -
\frac{1}{m}\big[V'(z)+U_{11}\rho' + \epsilon_{MF}' \nonumber \\
& & +\epsilon_{KE}'+\epsilon_{SP}'+\epsilon_{QP}'\big];\\
\mathrm{where} \nonumber \\
\label{eq:hydroEvol3} \epsilon_{MF1}' & = & U_{11} \rho
\big[\delta_2 f
f' + \delta_c f'(1-2 f) \big], \\
\label{eq:hydroEvol3b} \epsilon_{MF2}' & = & U_{11} \rho'
\big[1 + \delta_2 f^2 + 2 \delta_c f(1-f) \big], \\
\label{eq:hydroEvol5}\epsilon_{KE}' & = & (1-f)\bigg[\frac{1}{2}m
v_1^2 \bigg]' +
f\bigg[\frac{1}{2}m v_2^2 \bigg]' \nonumber \\
& & + f(1-f)\frac{\rho'}{\rho}\bigg[\frac{1}{2}m (v_1-v_2)^2\bigg] \nonumber \\
& & + f(1-f)\bigg[\frac{1}{2}m (v_1-v_2)^2\bigg]' \nonumber \\
& & - \frac{1}{2}m(v_1-v_2)f'\big[f v_1 + (1-f) v_2\big], \\
\label{eq:hydroEvol5}\epsilon_{SP}' & =&  \frac{\hbar^2}{4m}
\bigg[\frac{f'^2}{f(1-f)}\bigg]'+\frac{\hbar^2}{4m}\frac{\rho'}{\rho}\bigg[\frac{f'^2}{f(1-f)}\bigg], \nonumber \\
\label{eq:hydroEvol6}\epsilon_{QP}' & = & \frac{\hbar^2}{8m}
\bigg[\frac{\rho'^2}{\rho^2} - 2 \frac{\rho''}{\rho} \bigg].
\end{eqnarray}

We next linearize in the velocity field $v_c$ and density
fluctuations $\delta \rho = \rho- \rho_0$.  In this context,
$\rho_0$ is defined as the single component ($f=0$) stationary
solution:

\begin{eqnarray}
\label{eq:initRho} V+ U_{11}\rho_0+
\frac{\hbar^2}{8m}\frac{(\rho_0'^2 - 2 \rho_0
\rho_0'')}{\rho_0^2}=0.
\end{eqnarray}

\noindent Performing the linearization and eliminating $v_c$
yields a second order equation for the density fluctuations:

\begin{eqnarray}
\label{eq:linhydroEvol} \ddot{\delta\rho} & = &
\frac{\rho_0}{m}\bigg[U_{11} \delta\rho'+
\epsilon_{MF1}^{(1)'}+\epsilon_{MF2}^{(1)'}+\epsilon_{KE}^{(1)'}+\epsilon_{SP}^{(1)'}\bigg]'
\nonumber \\
& & \frac{\rho_0'}{m}\bigg[U_{11} \delta\rho'+
\epsilon_{MF1}^{(1)'}+\epsilon_{MF2}^{(1)'}+\epsilon_{KE}^{(1)'}+\epsilon_{SP}^{(1)'}\bigg]
\end{eqnarray}

\noindent where
$\epsilon_{MF1}^{(1)'},\epsilon_{MF2}^{(1)'},\epsilon_{KE}^{(1)'},$
and $\epsilon_{SP}^{(1)'}$ and are obtained by making the
replacement $\rho \rightarrow \rho_0$ in the corresponding
expressions (\ref{eq:hydroEvol3}-\ref{eq:hydroEvol5}).  In this
expression we have dropped terms involving the product of the
fluctuations $\delta \rho$ with kinetic energy terms and relative
scattering length differences $\sim\delta_c,\delta_2$.

Equation~(\ref{eq:linhydroEvol}) is quite widely applicable, however
it turns out that in practice we can simplify things further; in the
Thomas-Fermi regime the spatial derivatives of the background
density $\rho_0$ are small compared to the mean field and as a
result the second line of Eq.~(\ref{eq:linhydroEvol}) can be
neglected relative to the first line. The result is an intuitively
simple picture: The first term implies the density fluctuations obey
a phonon-like dispersion term with usual sound speed in the
condensate $c_0 \equiv \sqrt{U_{11} \rho_0/m}$, while the remaining
terms of the first line provide various source terms which seed
these density fluctuations and are non-zero only in locations where
there is a spin-excitation (that is, where $f \not= 0$).  These
source terms occur both because of differences in the mean field
interaction between the components as well as kinetic energy and
quantum pressure in the two wavefunctions. Because the spin dynamics
are much slower than the sound speed propagation, the source terms
appear to be nearly stationary to the phonons.

Armed with this picture, we can now interpret
Fig.~\ref{fig:breathEx}(c).  The initial spin excitation gives
non-zero source terms that generate phonons, which then propagate
through the BEC towards the boundary.  Meanwhile, in the region of
the spin excitation, the fluctuations are driven into a quasi-steady
state solution. In an infinite medium the phonons would continue,
however here they reflect off the boundaries and so repeatedly cross
the area of the spin excitation.   In the Fig.~\ref{fig:breathEx},
such a crossing occurs at 30~ms, while reflections off the
boundaries occur at 15~ms and 45~ms.

In many cases of interest, the mean field contribution
$\epsilon_{MF1}^{(1)'}$ dominates the kinetic energy contributions
in Eq.~(\ref{eq:linhydroEvol}).  In this case it is easy to solve
for the quasi-steady state solution $\delta\rho^{\mathrm{(ss)}}$ by
setting $\ddot{\delta\rho}=0$:

\begin{eqnarray}
\label{eq:deltaRhoSS} \delta\rho^{\mathrm{(ss)}} & = & -
\rho_0\bigg[\delta_c f + \frac{1}{2} f^2(\delta_2 - 2 \delta_c)
\bigg]
\end{eqnarray}

\noindent When $f \ll 1$ this reduces to $\delta\rho^{\mathrm{(ss)}}
= - \delta_c |\psi_2|^2$, which predicts fluctuations which are
directly proportional to the density in $\ket{2}$.  This approximate
solution holds fairly well in the example, as shown by the
dot-dashed curves in Fig.~\ref{fig:breathEx}(a).  At 15~ms and 45~ms
the sound waves are primarily at the BEC edge and this quasi-steady
state solution dominates in the area of the spin-excitation.  At
30~ms, the sound waves are passing through the spin excitation
region and are comparable to the quasi-steady state part.  These
sound waves end up having virtually no impact on the evolution of
$\psi_2$ due to the fact that their evolution is on a completely
different and independent time scale.  Stated another way, though
the sound waves overlap $\psi_2$ each time they cross the BEC, the
contributions from each crossing tend to be out of phase and their
net contribution to the evolution of $\psi_2$ washes out to nearly
zero. On the other hand the quasi-steady state part, given by
Eq.~(\ref{eq:deltaRhoSS}), can have a large impact on the evolution
of $\psi_2$.

Explicit numerical calculation of the various source terms reveals
that the mean field term indeed dominates by more than an order of
magnitude, meaning Eq.~(\ref{eq:deltaRhoSS}) should hold.
Furthermore, inspection of the figure shows that the initial peak
relative density is $f \sim 1/5$ and the first term should dominate.

\subsection{\label{subsec:EffectiveOneComp} Effective one-component GP
equation}

The existence of this simple solution allows us to then solve for
the much slower evolution of $\psi_2$.  Re-writing
(\ref{eq:GP2comp2}) in terms of $\delta_c,\delta_2$, eliminating
$|\psi_1|^2$ in favor of $\rho$ and using our solution of
(\ref{eq:initRho}) for the initial density profile $\rho_0$ allows
us to write:

\begin{eqnarray}
\label{eq:psi2Eff0}i \hbar \frac{\partial \psi_2}{\partial t} & =
& \bigg[\frac{-\hbar^2}{2m}\frac{\partial^2}{\partial z^2} -
\delta_c V(z) + U_{11}(\delta_2-\delta_c) |\psi_2|^2  \nonumber
\\ & & + U_{11} \delta \rho \bigg] \psi_2  \\
\label{eq:psi2Eff} & \approx &
\bigg[\frac{-\hbar^2}{2m}\frac{\partial^2}{\partial z^2} -
\delta_c V(z)  \nonumber
\\ & & + U_{11}(\delta_2-2 \delta_c) |\psi_2|^2 \bigg]
\psi_2.
\end{eqnarray}

\noindent where in the last line we have substituted our result for
the quasi-steady state $\delta\rho \approx - \delta_c |\psi_2|^2$.
This is just a one-component GP equation governing $\psi_2$ with the
trap and interaction coefficients renormalized by the
$\delta_2,\delta_c$ and is the central result of our paper. Note
that both the linear and nonlinear terms can have either sign,
meaning that the qualitative behavior of $\psi_2$ is quite sensitive
to the sign of the relative scattering lengths.  Note also that the
equation is consistent with our assumption that the $\psi_2$
dynamics are slower than the phonon dynamics when
$|\delta_2|,|\delta_c| \ll 1$.

\begin{figure}
\includegraphics[width=\columnwidth]{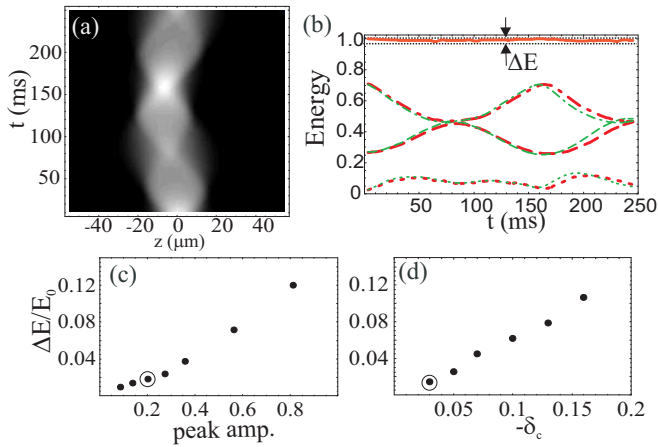}
\caption{\label{fig:oneCompMod}\textbf{(a)} Evolution of
$|\psi_2|^2$ according to the effective one-component GP equation
(\ref{eq:psi2Eff}).  \textbf{(b)} Contributions to the energies,
from Eq.~(\ref{eq:energies}), $E_k$ (dotted curves), $E_V$ (dashed),
$E_{\mathrm{int}}$ (dot-dashed) and total $E$ (solid) for the true
two-component evolution according to
Eqs.~(\ref{eq:GP2comp1}-\ref{eq:GP2comp2}) (thick, orange curves)
and effective one-component model Eq.~(\ref{eq:psi2Eff}) (thin,
green curves).  All energies are normalized by the initial total
energy $E_0=E(t)$. The dotted lines indicate the maximum and minimum
values of $E$ reached in the true two-component evolution. Their
difference defines $\Delta E$. \textbf{(c)} The relative
fluctuations of $E$ in the two-component model, normalized to the
initial energy $E_0$, with the initial peak amplitude $f$ varied.
The circled point indicates the case plotted in (a),(b).
\textbf{(d)} The relative energy fluctuations $\Delta E/E_0$ with
parameters the same as Fig.~\ref{fig:breathEx}, but varying
$-\delta_c$ and keeping $\delta_2=\delta_c$.}
\end{figure}

We made a number of assumptions in deriving Eq.~(\ref{eq:psi2Eff})
and so we now turn our attention to studying the quantitative
accuracy of this equation in several cases.   Simultaneously, this
will allow us to explore a variety of qualitative distinct regimes
it predicts.  Addressing first our example from
Fig.~\ref{fig:breathEx} we plot in Fig.~\ref{fig:oneCompMod}(a) the
evolution of the density profile $|\psi_2|^2$ as predicted by
Eq.~(\ref{eq:psi2Eff}).  One sees no visible deviations from the
full two component calculation in Fig.~\ref{fig:breathEx}(b). Note
that because $\delta_c=\delta_2=-0.03$ we get an effective repulsive
interaction coefficient $0.03 U_{11}$ and trapping potential $0.03
V(z)$ giving rise to the breathing and dipole motion observed in the
calculation.

A quantitative comparison can be made by calculating the energy
functionals:

\begin{eqnarray}
\label{eq:energies}
E_ K & \equiv & \int dz |\psi_2'|^2;  \nonumber \\
E_ V & \equiv & -\delta_c\int dz V(z)|\psi_2|^2;  \nonumber \\
E_{\mathrm{int}} & \equiv & \frac{1}{2}(\delta_2-2 \delta_c)\int dz |\psi_2|^4.  \nonumber \\
E & = & E_K+E_V+E_{\mathrm{int}}; \nonumber \\
\end{eqnarray}

\noindent which we plot in Fig.~\ref{fig:oneCompMod}(b) for both for
the effective one-component model (thin, green curves) and the full
two-component prediction (thick, orange curves).  One sees the
curves track each other very closely and, furthermore, that all
three contributions $E_K,E_V,E_{\mathrm{int}}$ are playing an
important role in the evolution, indicating the evolution of
$\psi_2$ is nonlinear in this case. The main deviation one observes
is a very slight difference in the time scale for the oscillatory
motion in the two cases. The total effective energy $E$ is
necessarily conserved for the one-component model, while this
quantity will only be conserved for the true two-component
calculation when Eq.~(\ref{eq:psi2Eff}) is providing an accurate
description of the evolution. Thus the fluctuations of this quantity
(the solid, orange curve) is a good measure of the validity of the
model and in the case there once sees these fluctuations $\Delta E$
are at a few percent level relative to the initial total energy
$E_0$.  In general, this error generally provides an estimate of the
error in the oscillatory time scale.

\section{\label{sec:applications} Applications of the model}

\subsection{\label{subsec:vectorSolitons} Prediction of breathing motion}

According to our model, Eq.~(\ref{eq:psi2Eff}), relative scattering
lengths in the regime in Fig.~(\ref{fig:breathEx}) ($\delta_c<0$ and
$\delta_2 - 2\delta_c > 0$) will give rise to an evolution of
$\psi_2$ analogous to a trapped interacting single component BEC. To
test this model and investigate the range of its applicability, in
Fig.~\ref{fig:oneCompMod}(c) we plot the effective energy
fluctuations for a number of cases with the same parameters as
Fig.~\ref{fig:oneCompMod}(a-b) but varying the initial peak relative
density $f=|\psi_2|^2/\rho$. One sees that even up to $f \approx
0.7$ the energy fluctuations are still less than 10 \%. In
Fig.~\ref{fig:oneCompMod}(d) we show a series varying the magnitudes
of $\delta_2,\delta_c$ (keeping $\delta_2 = \delta_c < 0$) and see
an approximate linear dependence.

We note that because this regime generally leads to smooth breathing
behavior, it may be well suited to performing controlled information
processing via the two-component dynamics.  In particular, the
linear evolution can be predicted by decomposing the wavefunction
into the harmonic oscillator states of the effective potential, and
amount of nonlinearity can be controlled by the number of atoms
coupled into $\ket{2}$.  Furthermore, much can be borrowed from the
vast existing literature on one-component BEC dynamics, including
analytic predictions for the ground states in the Thomas-Fermi
regime \cite{TF} and the spectrum of linear excitations from this
ground state \cite{oneCompExcitations}.

\subsection{\label{subsec:vectorSolitons} Phase separation and vector
solitons}

\begin{figure}
\includegraphics[width=\columnwidth]{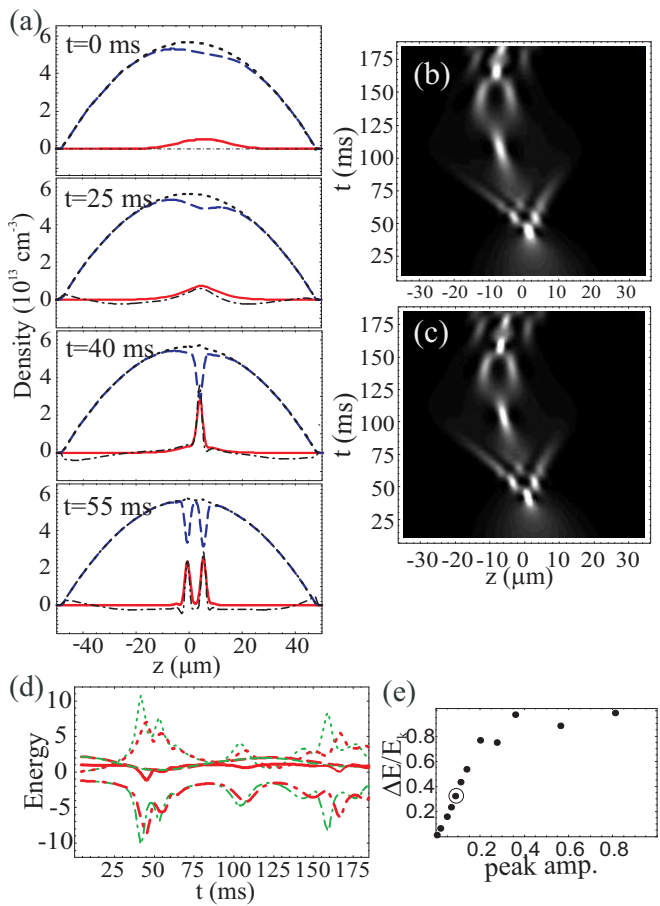}
\caption{\label{fig:SolEx}   Numerical simulations with
$\delta_c=-0.03, \delta_2=-0.09$, giving an effective attractive
interaction.  All conventions are the same as in
Figs.~\ref{fig:breathEx}-\ref{fig:oneCompMod} \textbf{(a)} Snapshots
of the two component's densities profiles and the density
fluctuations. \textbf{(b)} Time evolution of the density profile
$|\psi_2|^2$, \textbf{(c)} Effective one-component prediction
\textbf{(d)} Energies. \textbf{(e)} Total energy fluctuations
(normalized by the maximum value reached by $E_k$). The circled
point again indicates the case shown, and the other points indicate
simulations with the same parameters, but varying the initial peak
amplitude of $f$.}
\end{figure}

One particularly interesting behavior of two species BECs is
phase-separation \cite{phaseSepExp,phaseSepTh} which is predicted to
occur when $U_{12}^2 > U_{11} U_{22}$.  In this case it is
energetically favorable for the two components to separate into a
series of non-overlapping domains.  The manner in which this
separation takes place dynamically has not been addressed in detail.
According to our model Eq.~(\ref{eq:psi2Eff}), a species $\ket{2}$
contained in a condensate of another species $\ket{1}$ will
effectively act as an attractive BEC when $\delta_2-2 \delta_c<0$,
which, to first order in $|\delta_2|, |\delta_c|$ is equivalent to
the above phase separation criteria.  Such a case
($\delta_2=-0.09,\delta_2=-0.03$) is presented in
Figs.~\ref{fig:SolEx}(a-b).  One sees that, indeed, the $\ket{2}$
condensate acts as if it has attractive interactions which leads to
phase separation.  The wavefunction $\psi_2$ first collapses then
suddenly splits into two distinct soliton-like structures.
Fig.~\ref{fig:SolEx}(b) shows that this evolution continues, with
the the number of distinct structures alternating between 1,2 and 3.
The effective one-component model here allows us to map this
two-component problem onto the problem of soliton train formation in
single component, attractive interaction condensates, studied both
experimentally \cite{solitonTrainExp}  and theoretically
\cite{solitonTrainTh}, and thus acts as an intuitively simple and
quantitatively useful guide in predicting the formation and dynamics
of two-component (or vector) solitons \cite{twoCompSolitonsBEC}. The
effective one-component prediction is plotted in
Fig.~\ref{fig:SolEx}(c) and one sees the model gives a qualitatively
accurate prediction of the behavior.

One sees in Fig.~\ref{fig:SolEx}(d) that the quantitative error in
the evolution is slightly larger than in the effective repulsive
case with visible differences in the magnitude of the initial
kinetic and interaction energy oscillations and especially in the
magnitude and timing of later oscillations.  This is primarily
because the relative density $f$ grows to be quite large during the
phase separation.  The small $f$ prediction $\delta \rho=-\delta_c
|\psi_2|^2$ still provides a remarkably good estimate, as shown in
Fig.~\ref{fig:SolEx}(a).  While Eq.~(\ref{eq:deltaRhoSS}) would seem
to imply that including a next order nonlinearity term $\sim f^2$ in
the evolution could further improve the model, we found numerically
that this did not significantly improve the quantitative accuracy. A
further complication is that (the linearized versions of) the
kinetic energy and quantum pressure source terms
Eqs.~(\ref{eq:hydroEvol3a}-\ref{eq:hydroEvol6}) can be more
important (relative to the mean field term
Eqs.~(\ref{eq:linhydroEvol3})).  While explicit calculation in the
case shown reveals they were still smaller than the mean field term
by a factor $\sim 10$, this is enough to have some effect.
Fig.~\ref{fig:SolEx}(e) shows the energy fluctuation error (this
time relative to the peak kinetic energy $E_k^{(\mathrm{max})}$
since the initial total energy $E_0$, being the sum of two large
numbers of opposite sign, is near zero).  One sees an approximately
linear dependence of the energy fluctuations with the initial peak
$f$, which saturates when the initial peak value reaches about
$0.25$. For simulations with $f$ higher than the case plotted in
Figs.~\ref{fig:SolEx}(a-c), the number of solitons predicted was
incorrect.

\subsection{\label{subsec:repulsive} Effective repulsive potentials}

In the examples we have shown so far, we have chosen $\delta_c<0$,
since this leads to an effective trapping potential for $\psi_2$.
The model is accurate for the opposite case as well, however it
predicts (and we indeed observe) that $\psi_2$ is pushed to the edge
of the BEC due to the effective repulsive potential.  This occurs on
a time scale of about 20~ms for $\delta_c=0.03$ and the initial BEC
parameters used in this paper.  The model eventually breaks down
when $\psi_2$ reaches the BEC boundary, since the kinetic energy
becomes comparable to the mean field potential $U_{11} \rho_0$, and
the Thomas-Fermi assumption is no longer valid. Similar approaches
could be constructed to account for this boundary region in certain
cases (for example, in \cite{processing} this was done for a case
with a negligible nonlinearity term) but this is beyond our scope
here.

\subsection{\label{subsec:Rb87} Vanishing nonlinearity in ${}^{87}$Rb}

A particular case of interest in present day experiments is the
hyperfine levels $\ket{1}=\ket{F=1,M_F=-1}$ and
$\ket{2}=\ket{F=2,M_F=+1}$ of ${}^{87}$Rb.  These two levels are
approximately equally trapped magnetically and have a very small
inelastic collision rate \cite{inelastic}, allowing them to overlap
for very long times and maintain their coherence. Interestingly, the
scattering lengths in this case ($\delta_c=-0.03, \delta_2=-0.06$)
\cite{scatteringLengthRb} are such that the effective nonlinearity
vanishes and we get a very simple linear evolution of $\psi_2$ which
could be predicted by simply decomposing the initial state into the
eigenstates of the effective harmonic oscillator potential
$-\delta_c V(z)$. There will be higher order terms $\sim
|\delta_2|^2, |\delta_c|^2$ which eventually introduce nonlinearity
at longer times (and in fact this system is eventually phase
separating \cite{multipleComponentJILA}). However the linear model
could prove to be a powerful tool for predicting otherwise
computationally expensive two-component evolution in two and three
dimensions.

\subsection{\label{subsec:stat} Quasi-stationary solutions}

\begin{figure}
\includegraphics[width=\columnwidth]{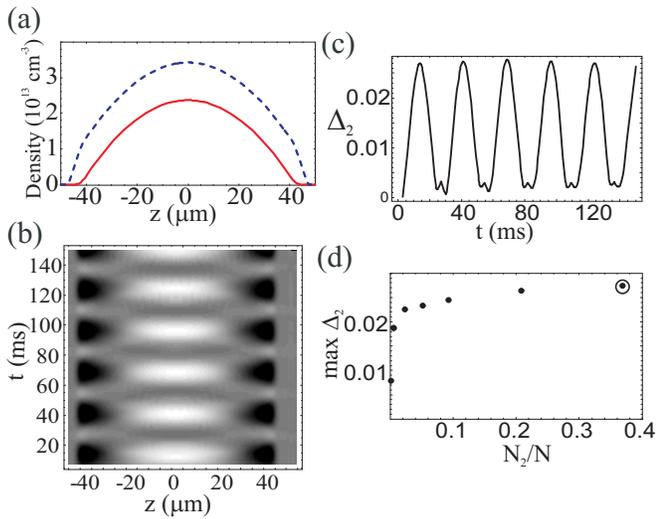}
\caption{\label{fig:stat}\textbf{(a)} The ground state density
profile $|\psi_2|^2$ obtained by finding the effective ground state
of Eq.~(\ref{eq:psi2Eff}) for a case with $N_2=0.38 N$, with
$|\psi_1|^2$ chosen such that the total density $\rho$ corresponds
to the original ground state profile $\rho_0$. \textbf{(b)} The
deviations from the initial density
$|\psi_2(z,t)|^2-|\psi_2(z,0)|^2$ are quasi-periodic. \textbf{(c)}
The deviation parameter $\Delta_2 = \int dz
(|\psi_2(z,t)|^2-|\psi(z,0)|^2)/\int dz |\psi_2(z,t)|^4$ as versus
time. \textbf{(d)} The maximum value reached by $\Delta_2$ as a
function of the fraction of atoms in $\ket{2}$, $ N_2/N$.}
\end{figure}

As a final application of our model, we consider the stationary
states of the effective one-component model.  In a case with only a
small density $|\psi_2|^2$ the eigenstates of the harmonic
oscillator potential will be stationary states of $\psi_2$.  The
ground state is simply a Gaussian wavefunction which could easily be
created with a stopped light pulse \cite{processing}.  Our model
would then predict the state written onto the wavefunction $\psi_2$
is then stationary and therefore quite robust with respect to
storage of information. We have indeed observed numerically that the
Gaussian ground state of the effective potential is stationary. Such
an approach was used in \cite{vortexStorage} to predict robust
storage of optical vortex states in BECs in three-dimensional
geometries.  In this work, the stability was checked with a full
calculation of the two-component evolution, but the one-component
model provided a way to construct an accurate estimate for a stable
configuration.

One of the motivations for stationary two-component states is the
possibility of atom-atom interaction induced spin-squeezing, as
proposed in \cite{spinSqueezing}.  In that proposal, for
${}^{23}$Na, $\delta_c=-0.08,\delta_2=0$. To achieve spin-squeezing,
one must have two components, each with a comparable density,
overlap for a considerable time.  Ideally, the mean-field dynamics
should be kept to a minimum to prevent them from washing out the
squeezing dynamics. This latter requirement could  be accomplished
by choosing the number of atoms in each component consistent with
the breathe-together solutions, whereby the relative density between
the two components ($f$) is constant across the BEC
\cite{dynamics,breatheTogether}, while the overall density profile
breathes. However, our model predicts quasi-stationary solutions for
any number of atoms in $\ket{2}$. Furthermore, if one wished to
write the squeezed state via slow light techniques, an inhomogenous
configuration, where $\psi_2$ is contained in $\psi_1$ and vanishes
at the BEC boundaries, would be essential to prevent spontaneous
emission events. To our knowledge, two species BEC stationary
states, with an inhomogenous relative density profile $f$, have not
been previously predicted or experimentally observed.

As an example, in Fig.~\ref{fig:stat}(a) we show a ground state of
the effective one-component GP equation (\ref{eq:psi2Eff}),
obtained by propagating the equation in imaginary time, and
holding the number of atoms in $\ket{2}$ $N_2 = N \int dz
|\psi_2|^2$ constant.  In this case, the nonlinear term
$E_{\mathrm{int}}$ is quite significant (it dominates the kinetic
energy $E_k$ in the ground state).  The density $|\psi_1|^2$ is
then chosen so that the total density $\rho_0$ corresponds the
ground state of a pure $\ket{1}$ condensate $\rho_0$.

A time-dependent evolution of the full two-component
Eqs.~(\ref{eq:GP2comp1}-\ref{eq:GP2comp2}) with this initial state
then reveals that indeed this configuration is nearly stationary.
The only observed motion is a small in-phase breathing of both
components, with a magnitude governed by $\delta_c$.
Figure~\ref{fig:stat}(b) shows the variations of the density
$|\psi_2|^2$ from its initial value as a function of time, which
exhibits this breathing motion. Thus, not only is the motion of
$\psi_2$ small, but the small deviations are approximately periodic
and so $|\psi_2|^2$ remains near its initial value for very long
times. The parameter $\Delta_2$, plotted in Fig.~\ref{fig:stat}(c)
and defined in the caption, characterizes the total deviation and is
seen to be nearly periodic with a maximum amplitude of about 0.025.

We performed similar simulations for a variety of values $N_2$ and
plotted the maximum $\Delta_2$ reached in each case.  The results
are presented in Fig.~\ref{fig:stat}(d).  In the limit of small
$N_2$, where the nonlinearity is negligible the effective ground
state is truly stationary.  As the nonlinearity becomes more
important, the ground state grows due to effective repulsive
interactions, the small breathing motion occurs and we get the
deviations $\Delta_2$. The magnitude of $\Delta_2$ saturates at
around 0.025 (see Fig.~\ref{fig:stat}(d)). The saturation occurs at
the point the Thomas-Fermi solution \cite{TF} of the effective
one-component model becomes accurate. The saturation value should
scale roughly linearly with $|\delta_c|$.

In the regime $N_2 \sim N_1$, the effective ground state
eventually becomes larger than the original condensate, in which
case the present approach fails. However, in that limit, the
solutions smoothly cross over into the breathe-together solutions
discussed in \cite{dynamics,breatheTogether}.

\section{\label{sec:summary} Summary}

In summary, we have derived an equation of motion for the total
density fluctuations in dynamic two-component condensates. We have
then used this to derive an effective one-component GP equation
(\ref{eq:psi2Eff}) for the smaller component, with effective
potentials and interaction coefficients which depend in a simple way
on the relative scattering lengths in the system.   We have studied
and tested the predictions of this model in both effective repulsive
and attractive interactions.  Our model provides an intuitively
simple way to predict the dynamics of two component BECs and allow
us to make correspondences to results already obtained for
one-component BECs. In particular, our model provides new insight on
the dynamics of phase separation and the formation of vector
solitons. We noted that magnetically trapped ${}^{87}$Rb provides a
particularly interesting example in which the motion is governed by
a simple linear Schroedinger equation, allowing simple analytic
predictions of motion in cases that would otherwise require
solutions of the underlying nonlinear Schroedinger equations.
Finally, we have applied it to predict the existence of
quasi-stationary two-component configurations. These solutions
promise to be  useful for applications involving information storage
using stopped light as well as inducing spin squeezing in these
systems. A combination of these two techniques may provide a new
technique to generate squeezed light.


\begin{thebibliography}{99}

\bibitem{BEC}  M. Inguscio, S. Stringari, and C.
Wieman, eds., {\it Bose-Einstein Condensates in Atomic Gases,
Proceedings of the International School of Physics Enrico Fermi,
Course CXL}, (International Organisations Services B.V.,
Amsterdam, 1999).

\bibitem{phaseSepExp} H.-J. Miesner {\it et al.}, Phys. Rev. Lett.
{\bf 82}, 2228 (1999).

\bibitem{phaseSepTh} T-L. Ho and V.B. Shenoy, Phys. Rev. Lett.
{\bf 77}, 3297 (1996); H. Pu and N.P. Bigelow, Phys. Rev. Lett.
{\bf 80}, 1130 and 1134 (1998); E. Timmermans, Phys. Rev. Lett.
{\bf 81} 5718 (1998); P. Ao and S.T. Chui, Phys. Rev. A {\bf 58},
4836 (1999).

\bibitem{twoCompSolitonsBEC} Th. Busch and J.R. Anglin, Phys. Rev. Lett. {\bf 87}, 101401
(2001).

\bibitem{twoCompVortex} M.R. Matthews, {\it et al.}, Phys. Rev.
Lett., {\bf 83} 2498 (1999).

\bibitem{twoCompVortexStab} J.J. Garcia-Ripoll and V.M.
Perez-Garcia, Phys. Rev. Lett. {\bf 84}, 4264 (2000).

\bibitem{twoCompVortexLatt} E.J. Mueller and T.-L. Ho, \prl {\bf 88}, 180403 (2002);
K. Kasamatsu, {\it et al.}, {\it ibid.} {\bf 91}, 150406 (2003).

\bibitem{vortexStorage} Z. Dutton and J. Ruostekoski, Phys. Rev.
Lett. {\it in press}, cond-mat/0405159.

\bibitem{spinWaves} T. Nikuni and J.E. Williams, J. Low Temp.
Phys. {\bf 133}, 323 (2003).

\bibitem{dynamics} A. Sinatra and Y. Castin, European Phys.
Jour. D {\bf 8}, 319 (2000).

\bibitem{breatheTogether} S.D. Jenkins and T.A.B. Kennedy, Phys. Rev. A {\bf 68}, 053607
(2003).

\bibitem{multipleComponentJILA} D. S. Hall, M. R. Matthews, C. E. Wieman, and E. A.
Cornell, Phys. Rev. Lett. {\bf 81}, 1539 (1998); {\it ibid.} {\bf
81}, 1543 (1998).

\bibitem{symBreaking} B.D. Esry and C.H. Greene, Phys. Rev. A {\bf
59}, 1457 (1999).

\bibitem{twoCompStab} C.K. Law, H. Pu, N.P. Bigelow, and J.H.
Eberly, Phys. Rev. Lett. {\bf 79}, 3105 (1997).

\bibitem{processing} Z. Dutton and L.V. Hau, Phys. Rev. A, {\it in press}; quant-ph/0404018.

\bibitem{stoppedLight} C. Liu, Z. Dutton, C.H. Behroozi, and L.V. Hau, Nature {\bf
409}, 490 (2001). D.F. Phillips, A. Fleischhauer, A. Mair, R.L.
Walsworth, and M.D. Lukin, Phys. Rev. Lett. {\bf 86}, 783 (2001).

\bibitem{feshbach} S. Inouye {\it et al.}, Nature {\bf 392},
151 (1998);  S.L. Cornish {\it et al.}, Phys. Rev. Lett. {\bf 85},
1795-1798 (2000).

\bibitem{oneCompExcitations} S. Stringari, Phys. Rev.
Lett. {\bf 77}, 2360 (1996).

\bibitem{solitonTrainExp} K.E. Strecker, G.B. Partridge, A.G. Truscott, and R.G. Hulet, Nature {\bf 427}, 150 (2002);
L. Khaykovich, {\it et al.}, Science {\bf 296}, 1290 (2002).

\bibitem{solitonTrainTh} U. Al Khawaja, {\it et al.} Phys. Rev.
Lett. {\bf 89}, 200404 (2002); L. Salasnich, A. Parola and L.
Reatto {\bf 91}, 080405 (2003); L.D. Carr and J.Brand, Phys. Rev.
Lett. {\bf 92}, 040401 (2004).

\bibitem{spinSqueezing} A. Sorensen, L.-M. Duan, J.I.
Cirac, and P. Zoller, Nature {\bf 409}, 63 (2001).

\bibitem{opticalTrap} D.~M. Stamper-Kurn, {\it et
al.}, Phys. Rev. Lett. {\bf 80}, 2027 (1998).

\bibitem{sound} M.~R. Andrews, {\it et
al.}, Phys. Rev. Lett. {\bf 79}, 553 (1997).

\bibitem{TF} G. Baym and C.J. Pethick, Phys. Rev. Lett., Phys. Rev.
Lett. {\bf 76}, 6 (1996).

\bibitem{inelastic} C.J. Myatt,  E.A. Burt, R.W. Ghrist, E.A. Cornell,
and C.E. Weiman, Phys. Rev. Lett. {\bf 78}, 586 (1997); J.P.
Burke, C.H. Greene, J.L Bohn, Phys. Rev. Lett. {\bf 81}, 3355
(1998). A Mathematica notebook is available at
http://fermion.colorado.edu/$\sim$chg/Collisions/.

\bibitem{scatteringLengthRb} J.M. Vogels, R.S. Freeland, C.C. Tsai,
B.J. Verhaar, and D.J. Heinzen, Phys. Rev. A {\bf 61} 043407
(2002).

\end{thebibliography}
\end{document}